\documentstyle[psfig]{aa}

\newcommand{\etal }{{\it et al.}}

\input{epsf}

\begin{document}

\thesaurus{11.07.1;11.09.5;11.09.4;11.19.3}

\title
{CO in blue compact and star burst galaxies}

\author{
P.M. Gondhalekar\inst{1} \and
L.E.B. Johansson\inst{2} \and
N. Brosch\inst{3} \and I.S. Glass\inst{4} \and
E. Brinks\inst{5} 
}

\date{}

\institute {Rutherford Appleton Laboratory, Chilton, OXON, OX11
0QX, England. \and
Chalmers University of Technology, Onsala Space Observatory,
S-439 92 ONSALA, Sweden. \and
Department of Physics \& Astronomy, Tel Aviv University, Tel Aviv,
Israel. \and
South African Astronomical Observatory, PO Box 9, Observatory 7935,
Cape Town, South Africa. \and
Departamento de Astronomia,
Universidad de Guanajuato, Apdo. Postal 144, Guanajuato, Mexico.
}
\offprints{P M Gondhalekar}
\date{received 1900; accepted 2000}

\maketitle

\begin{abstract}

$^{12}$CO(J=1$\rightarrow$0) observations of 34 blue compact and star burst galaxies
are presented.  
Although these galaxies are experiencing vigorous star formation 
at the current epoch, CO has been detected in only five of them. 
The five detections
reported in this paper are all in galaxies with relatively red
colours, (B-V)$_{0}>$0.4.

The new observations, when combined with previously
published data on CO in BCGs, indicate that CO luminosity
decreases with  absolute luminosity of BCGs. Since 
the absolute luminosity of a galaxy is correlated with its metallicity,
these results confirm that low 
metallicity BCGs have low abundances of CO gas. We also show
that the star formation rate determined from the H$_{\beta}$\
luminosity is lower than that determined from the far infrared luminosity.

\end{abstract}

\section {Introduction}
Blue Compact Galaxies (BCGs) are the smallest star-forming extragalactic
objects and they are
experiencing massive bursts of star formation at the current epoch. 
The dwarf BCGs  lack the elaborate gas dynamics and spiral arms
which trigger star formation in giant galaxies and the mechanism for
star formation in these galaxies is still unidentified. An acceptable 
description of their nature and their evolution has also not yet emerged.
The metal and dust deficiency of BCGs suggests that they are either
young galaxies which have recently formed out of protogalactic gas clouds
or that they experience intermittent bursts of star formation
followed by periods of quiescence (Searle, Sargent \& Bagnuolo 1972,
Huchra 1977, Thuan et al. 1983, Gondhalekar et al. 1983,1986)

Observations of CO in nearby galaxies (Talbot 1980, Scoville \&
Young 1983) suggest that the formation of stars depends primarily on the 
availability of molecular gas. 
Thus observations of molecular gas in BCGs are essential to both
the understanding of star-formation within them as 
well as their overall evolution. CO observations of BCGs 
have been reported 
by Young et al. (1986), Israel \& Burton (1986), Tacconi \& Young (1987), 
Arnault et al. (1988), Sage et al. (1992) and  Israel, Tacconi \& Baas
(1995). Various criteria 
were used to select the samples of galaxies observed in these studies.  
The consensus seems 
to be that {\em metal-poor galaxies are deficient in CO}. The current 
programme was initiated to explore the validity of this
conclusion.  A sample of galaxies, covering a
large range in absolute magnitude and  60$\mu$m luminosity, 
has been observed in order to determine the CO
content in the galaxies experiencing different
levels of star bursts and spanning a range of metallicities. We should like
to emphasise that we are unable to determine the molecular content of
these galaxies from the CO observations as the CO-to-H$_{2}$\ conversion
factor is not known for the galaxies in this sample as the metallicity of
these galaxies is not known and the conversion factor is a function of
metallicity of a galaxy.

The sample of galaxies, the CO observations and the data 
reduction are described in section 2. In section 3
we investigate the relation between star-formation and CO in these galaxies . The 
conclusions are given in section 4.

\section {Sample, observations and data reduction}

For this study a sample of galaxies covering a large range in 
luminosity and therefore, metallicity, was selected from the list of 
Salzer, MacAlpine and Boroson (1989). We have selected 13 galaxies
classified as {\em dwarf HII hotspot galaxies, DHIIH,} by these authors.
To this we have added a
sample of nine {\em star burst nuclei, SBN}, (Salzer, MacAlpine and Boroson, 
1989). The galaxies in both samples have  H$_{\beta}$\ 
luminosity greater than $10^{38}$\ erg s$^{-1}$\ 
(H$_{0}$=75 km s$^{-1}$\ Mpc$^{-1}$\ and q$_{0}$=0.5 have been
used through out this paper) and all are
experiencing vigorous star bursts at the current epoch. 
Additional star burst galaxies, mainly from MacAlpine, Smith \& Lewis
(1977a,b) and MacAlpine \& Lewis (1978) were added to
this list. The full list of galaxies observed is given in Table 1. In 
total, 34 galaxies were observed. The sample covers about 
6$^{m}$\ in absolute luminosity. This is not a complete sample, it is,
however, a large sample and the number of BCGs is large enough to be 
representative as a class.

The observations of the $^{12}$CO(J=1$\rightarrow$0) (115.271 
GHz) line were performed
at the Onsala Space Observatory (OSO) 20m mm-wave telescope, 
the half-power beamwidth of which is 35$\arcsec$ at this
frequency. The observations
were made during three runs between February 1995 and March 1997.
A SIS mixer was used, tuned to the single side-band mode. Typical system
temperatures, corrected for rearward \\
spillover and atmospheric 
attenuation, were between 500 K and 1500 K depending on weather
conditions and galaxy redshifts. The back-end was a multichannel
receiver with a resolution of 1 MHz and a total bandwidth of 512 MHz
(corresponding to a velocity coverage of about 1300 km s$^{-1}$\ at 115 GHz)
Antenna pointing was about 3$\arcsec$\ rms on each axis and the main-beam
efficiency was $\eta_{mb}$=0.5 during all three observing runs. The observed
intensities, T$^{\ast}_{a}$, were ``chopper-wheel" calibrated and are related 
to the main-beam brightness temperature  T$_{mb}$\ by 
T$^{\ast}_{a}$=$\eta_{mb}\times$ T$_{mb}$. 

\begin{figure*}[!ht]
\centering
\psfig{figure=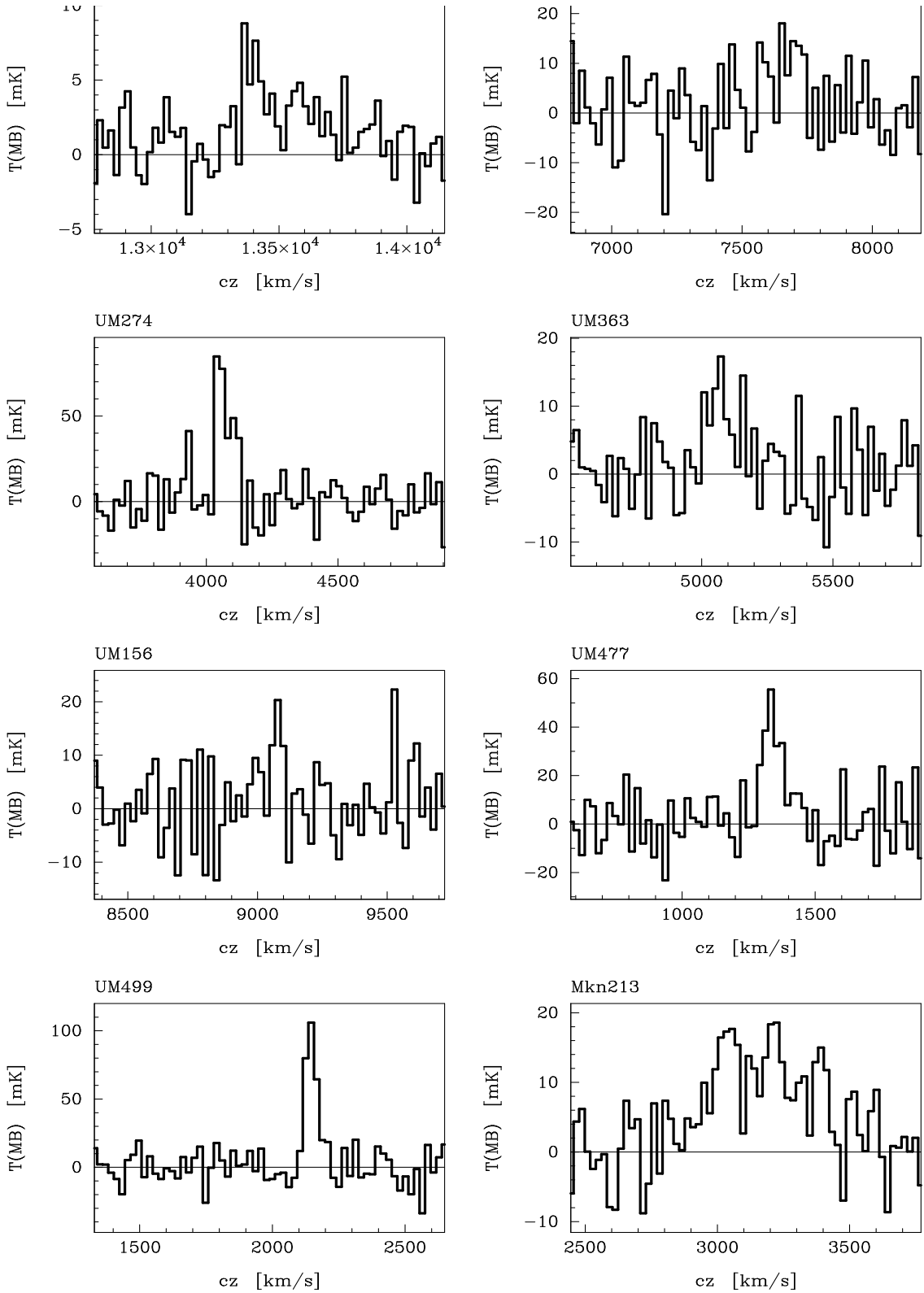,width=15.0cm,height=21.0cm}
\caption[]{CO spectra of eight star burst galaxies in the current sample. The intensity
scales are in T$_{mb}$\ (mK). The velocities are {\it 
cz} (km s$^{-1}$) and have not been corrected for the solar motion or the 
Virgo-centric motion.
The spectra have been smoothed to 20 km s$^{-1}$. Note that the
detections of UM191, UM363 and UM156 are only tentative.}
\end{figure*}

The detections and the upper limits of CO for all galaxies are given in 
Table 1, where the rms error (Col.7) is calculated for a resolution of 
20 km s$^{-1}$. This resolution seems to give the
best signal-to-noise for the detected line widths. 
The intensity scale is the main 
beam brightness temperature T$_{mb}$. The integrated CO
intensity $I_{CO} = \int T_{mb} dV$\ is given in Col. (11).
The 1$\sigma$\ error in the integrated line intensity, where the line has
been detected, is given in Col.
(12) and was calculated using Eq.(1) of Elfhag et al. (1996). These
errors depend on the inherent rms noise in the profile and the
uncertainties in baseline corrections. However, for most of the detected
lines the baseline range is small and only the rms noise defines the error.
This is, however,  not true in cases where only upper limits can be
determined, because in such cases the full spectrum defines the baseline
range and the curvature of the baseline is included in computing the
upper limits.

Of the 34 galaxies observed, CO was detected in five and there may be 
possible detection in three other cases. 
These eight spectra are shown in Fig. 1. In the following discussion the 
possible detections are regarded as upper limits.

Some galaxies in this sample (see Table 1) are AGNs. Only upper
limits have been obtained for CO in these galaxies. These
limits have been included in the following analysis but they 
do not affect our conclusions.

\setlength{\tabcolsep}{5pt}
\begin{table*}
\caption[]{$^{12}$CO in bluse compact and star burst galaxies}
\begin{flushleft}
\begin{tabular}{lllrrrrrrcrrcccc}
\hline
~~~ID &  
      & 
      & 
\multicolumn{3}{c}{RA(1950)} & 
\multicolumn{3}{c}{Dec(1950)} &
(B-V)$_{cor}$ & 
T$_{mb}$ &
$\sigma_{mb}$   &  
c$\times$z & 
FWHW         & 
I$_{CO}$ &
$\sigma_{CO}$   \\ 
      &  
      & 
      & 
      & 
      & 
      &  
      & 
      & 
      &
      &
mK    &
mK    &
km s$^{-1}$ & 
km s$^{-1}$ & 
K km s$^{-1}$  &
  \\
~~~~1 & 
  ~~2 &
~~3   & 
\multicolumn{3}{c}{4} & 
\multicolumn{3}{c}{5} & 
6     & 
7     & 
8~~   &
9     & 
10    & 
11    & 
12 \\
\hline
UM274&    &Sb   & 0&43&32&$-$1&59&41&    &76&12.&4065& 65&5.4       &0.7\\
UM286&    &     & 0&49&26&$-$0&45&31&0.53&  &13.&    &   &$<$0.8    &   \\
UM307&SBd &SBN  & 1&08&56&   1&03&24&0.41&  &4.8&    &   &$<$0.3    &   \\
UM323&    &DHIIH& 1&24&12&$-$0&54&15&0.40&  &11.&    &   &$<$0.6    &   \\
UM334&    &DHIIH& 1&30&05&$-$1&54&15&0.33&  &6.2&    &   &$<$0.4    &   \\
UM343&SBb &SBN  & 1&33&25&   0&24&29&0.78&  &8.6&    &   &$<$0.5    &   \\
UM351&    &DHIIH& 1&35&48&   1&38&48&0.56&  &4.9&    &   &$<$0.3    &   \\
UM363&SO-a&Sy 2 & 1&41&22&   2&05&56&1.03&13&4.3&5070&125&1.7$^\ast$&0.3 \\
UM374&    &DHIIH& 1&50&20&$-$1&08&52&0.60&  &8.1&    &   &$<$0.5    &    \\
UM385&    &Sy 1 & 1&57&16&   0&09&09&    &  &6.8&    &   &$<$0.3    &    \\
UM388&    &SBN  & 1&58&18&$-$1&46&50&0.72&  &4.6&    &   &$<$0.3    &    \\
UM393&SO  &Sy 1 & 2&03&43&$-$0&31&46&    &  &5.7&    &   &$<$0.3    &    \\
UM413&    &SBN  & 2&12&21&   2&00&47&0.75&  &3.4&    &   &$<$0.2    &    \\
UM418&SBb &SBN  & 2&17&07&$-$0&29&09&    &  &6.8&    &   &$<$0.4    &    \\
IIZw40&Sbc&     & 5&53&05&   3&23&05&0.82&23&6.6& 645& 45&1.1$^\ast$&0.5 \\
IZw18 &   &     & 9&30&30&  55&27&49&0.10&  &4.8&    &   &$<$0.3    &    \\
UM439 &   &DHIIH&11&34&02&   1&05&38&0.28&  &8.3&    &   &$<$0.5    &    \\
UM444 &   &DHIIH&11&37&38&$-$0&08&03&0.65&  &4.9&    &   &$<$0.3    &    \\
UM452 &SO &DHIIH&11&44&26&$-$0&00&57&0.64&  &13.&    &   &$<$0.7    &    \\
UM454 &   &DHIIH&11&45&43&$-$1&21&43&0.43&  &13.&    &   &$<$0.8    &    \\
UM456 &   &DHIIH&11&48&01&$-$0&17&23&0.35&  &11.&    &   &$<$0.7    &    \\
UM462 &   &DHIIH&11&50&03&$-$2&11&26&0.24&  &7.4&    &   &$<$0.5    &    \\
UM471 &   &DHIIH&11&58&56&$-$1&09&28&0.22&  &10.&    &   &$<$0.5    &    \\
UM477 &SBc&SBN  &12&05&36&   3&09&21&0.62&50&8.6&1340& 85&4.5       & 0.6   \\
UM483 &   &DHIIH&12&09&41&   0&21&00&0.02&  &14.&    &   &$<$0.8    &     \\
UM491 &   &DHIIH&12&17&18&   2&03&02&0.42&  &14.&    &   &$<$0.8    &     \\
UM499 &SO-a&SBN &12&23&09&   0&50&57&0.64&110&10.&2145&55& 6.2      & 0.5  \\
Mark213&SBa&    &12&29&01&  58&14&20&0.48&15 &4.6&3160&360&5.4      & 0.5  \\  
Mark54&Sc  &    &12&54&32&  32&43&07&0.27& 4 &1.8&13500&430&1.5     & 0.2  \\
UM530 &    &SBN &12&55&35&   2&07&54&0.51&   &5.8&     &   &$<$0.3  &      \\
UM549 &   &DHIIH&13&11&58&   2&49&44&0.39&   &5.2&     &   &$<$0.3  &      \\
UM641 &Sc &SBN  &14&07&21&$-$1&00&15&0.43&   &5.0&     &   &$<$0.3  &      \\
UM156 &Sb &Sy 1 &23&16&23&$-$0&01&50&    &15 &6.0& 9080& 65&1.0$^\ast$& 0.4  \\
UM191 &Im &     &23&54&26&$-$2&21&44&    & 13&5.7& 7675&150&2.0$^\ast$ & 0.4 \\
\hline
Column 2: & \multicolumn{15}{l}{ Galaxy type; LEDA catalogue 
(Paturel et al. 1997)} \\
Column 3: & \multicolumn{15}{l}{Galaxy type; Salzer et al. (1989)} \\
(B-V)$_{cor}$ & \multicolumn{15}{l}{ From Salzer et al. (1989), where
available, and from the LEDA catalogue otherwise} \\
Column 11: &   \multicolumn{15}{l}{An asterix denotes a possible detection 
only} \\ 
\hline
\end{tabular}
\end{flushleft}
\end{table*}

\begin{table*}[!ht]
\caption[]{CO and FIR luminosities and the HI mass in galaxies with CO
detection}
\begin{flushleft}
\begin{tabular}{lcccccrccr}
\hline
~~~ID  & m$_{B}$ & M$_{B}$ & z & log(L$_{H_{\beta}}$) & log(L$_{FIR}$) &
C$_{FIR}$  & log(L$_{CO}$) & log(M$_{HI}$) &D$_{25}$ \\
       &         &         &   & erg s$^{-1}$ & L$_{\sun}$ &  & K km
s$^{-1}$\ pc$^{2}$ & M$_{\sun}$ & $\arcsec$ \\
~~~1  &    2   &   3     &4      &5     &6     &7~~~~&   8   &9~~  &10  \\
\hline
 Mark54& 15.29&$-$21.81&0.0447&    &10.69&$-$0.289&9.70&     &40 \\
 UM274& 12.81&$-$21.22&0.1412&41.29&12.30& 0.363&9.16&11.61&75 \\
 UM477& 11.58&$-$19.85&0.0042&42.06& 9.46& 0.263&8.10& 9.90&256  \\
 UM499& 13.16&$-$19.40&0.0067&41.74& 9.71& 0.119&8.65& 8.92&128 \\
 Mark213&13.10&$-$20.96&0.0104&39.86&10.05& 0.242&8.95& 9.31&97 \\
       &     &      &      &     &     &      &    &     &   \\
$^{T}$NGC1569&10.68&$-$20.15&$-$0.0005&     & 8.51& 0.039&5.25 &8.04 \\
$^{T}$NGC4214&10.13&$-$17.32&0.0010&     &8.61&0.240&5.96&9.05 \\
$^{T}$NGC5253&10.77&$-$16.23&0.0010&41.16&8.86&$-$0.027&5.83&8.15 \\
UM465&  13.96&$-$17.01&0.0034&41.19&8.45&0.115&6.28&7.59 \\
NGC6822& 9.32&$-$17.55&0.0001&     &6.52& 0.169&4.48&7.90 \\
He2$-$10&12.46&$-$18.78&0.0022&   &9.44& 0.033&7.56&8.70 \\
$^{S}$IIZw40&15.48&$-$16.15&0.0026&40.75&8.99&$-$0.056&6.81&8.37 \\
$^{S}$Haro2& 12.97&$-$19.07&0.0051&     &9.45& 0.045&7.36&8.68 \\
$^{S}$Haro3& 13.21&$-$18.19&0.0034&     &8.75& 0.742&7.52&8.75 \\
$^{S}$Mark900&13.98&$-$17.50&0.0038&     &    &      & 6.30&8.93  \\
$^{S}$Mark86& 12.19&$-$16.70&0.0015&     &8.31& 0.298&6.18&8.28 \\
$^{S}$UM456& 15.21&$-$16.72&0.0056&41.09&8.28& 0.260&7.32&8.52 \\
$^{S}$UM462& 14.56&$-$16.33&0.0034&40.52&8.42& 0.030&7.51&8.15 \\
$^{S}$Mark297&13.44&$-$21.19&0.0157&    &10.63&0.147&9.53&10.06 \\
$^{S}$UM448& 14.39&$-$20.39&0.0182&42.04&10.49&0.114&9.33&9.67  \\
\hline
column 7 & \multicolumn{9}{l}{FIR colour, C$_{FIR}$=log($f_{100}/f_{60}$)} \\
$^{T}$         & \multicolumn{9}{l}{Data from Taylor, Kobulnicky \&
Skillman  (priv. com.)} \\
$^{S}$   &\multicolumn{9}{l}{Data from Sage et al. 1992} \\
UM465 & \multicolumn{9}{l}{Data from Taylor \etal\ (1995)} \\
NGC6822 & \multicolumn{9}{l}{Data from Davies (1972)} \\
He2$-$10 & \multicolumn{9}{l}{Data from Kobulnicky \etal\  (1995)} \\
\hline
\end{tabular}
\end{flushleft}
\end{table*}

\subsection {Comparison with previous observations}

Tacconi \& Young (1987) and Sage et al. (1992) have previously observed 
this CO line in II Zw 40. A comparison of the
measurements, taking into account the different beam areas, 
suggests that the present observations are 0.21 dex
higher than those of Tacconi \& Young (1987) and 0.35 dex higher than
those of Sage et al. (1992). However, the centroid velocity of the
present observations is $\approx$645 km s$^{-1}$\ while Tacconi \& Young
measure a velocity of $\approx$850 km s$^{-1}$\ and Sage et al. a value of 
$\approx$770 km s$^{-1}$\ respectively. Brinks \& Klein (1988) have
mapped II Zw 40 in HI with the VLA and they measured velocities 
between 750 and 800 km s$^{-1}$ . The poor agreement between the velocity
derived from our CO observations and the velocities seen in 
the VLA maps suggests a false detection of CO in the current observations.

Similarly, Taylor, Kobulnicky \& Skillman (priv. comm.) have reported CO
observations of UM477. The luminosity derived from the current 
observations is 0.42 dex higher than that observed by Taylor, Kobulnicky 
\& Skillman, but the velocity from the current observations 
is in good agreement with the velocity of 1318 km s$^{-1}$\
obtained by these authors.

\begin{table*}[!ht]
\caption[]{Upper limits of CO luminosities}
\begin{flushleft}
\begin{tabular}{lccccrrc}
\hline
~~~ID & m$_{B}$ & M$_{B}$ & z & log(L$_{H_{\beta}}$) & log(L$_{FIR}$) &
C$_{FIR}$ & log(L$_{CO}$) \\
      &         &        &   & erg s$^{-1}$ & L$_{\sun}$ &
       & K km s$^{-1}$\ pc$^{2}$ \\
~~~~1 & 2 &    3  &   4  &  5  &  6  &  7 & 8    \\
\hline
 UM286&15.2&$-$16.68&0.0037&     &     &    &7.20 \\
 UM323&16.1&$-$17.22&0.0064&40.34& 8.49&0.37&7.55  \\
 UM334&17.2&$-$16.98&0.0163&39.98& 9.51&0.12&8.19  \\
 UM351&18.2&$-$16.91&0.0250&40.53& 9.62&    & 8.43  \\
 UM363&13.6&$-$20.96&0.0170&41.69& 9.90&$-$0.005&8.85 \\
 UM374&17.6&$-$17.32&0.0191&40.87& 9.48&0.36&8.42  \\
 UM385&15.7&$-$23.30&0.1629&     &12.43&    &10.06 \\
 UM393&14.4&$-$22.22&0.0424&     &10.14&    &8.89  \\
 UM418&14.2&$-$21.27&0.0253&     &10.54&    &8.57  \\
 IZw18&15.6&$-$15.18&0.0025&     &     &    &6.44  \\
 UM439&15.3&$-$16.45&0.0037&39.99& 8.49&0.49&7.00  \\
 UM444&16.7&$-$19.00&0.0219&41.50& 9.56&    &8.32  \\
 UM452&15.5&$-$16.45&0.0046&39.35& 8.53&0.80&7.33  \\
 UM454&16.5&$-$17.68&0.0126&40.03& 9.17&0.45&8.26  \\
 UM456&15.5&$-$16.72&0.0056&41.09& 8.43&0.26&7.50  \\
 UM462&14.6&$-$16.33&0.0034&40.52& 8.59&0.03&6.92  \\
 UM471&18.2&$-$17.98&0.0351&41.07& 9.97&    &8.95  \\
 UM483&15.9&$-$17.12&0.0075&40.54& 8.41&    &7.81  \\
 UM491&15.8&$-$16.39&0.0063&42.09& 8.51&0.42&7.66  \\
 UM549&16.6&$-$18.08&0.0193&41.03& 9.23&    &8.21  \\
 UM307&14.4&$-$21.07&0.0228&42.05&10.48&0.32&8.35  \\
 UM343&14.1&$-$20.78&0.0173&41.83&10.25&0.26&8.34  \\
 UM388&17.7&$-$21.63&0.1243&41.78&11.36&0.62&9.83  \\
 UM413&17.6&$-$20.97&0.0263&40.81&11.17&0.30&8.30  \\
 UM530&16.8&$-$20.93&0.0665&42.09&11.06&0.12&9.28  \\
 UM641&15.2&$-$20.33&0.0240&42.28&10.00&0.26&8.40  \\
 UM156&13.5&$-$22.43&0.0293&42.02&10.34&0.41&9.12  \\
 UM191&15.3&$-$19.87&0.2451&     &     &    &9.27  \\
\hline
column 7 & \multicolumn{7}{l}{FIR colour, C$_{FIR}$=log($f_{100}/f_{60}$)} \\
column 8 & \multicolumn{7}{l}{Upper limit of CO luminosity } \\
\hline
\end{tabular}
\end{flushleft}
\end{table*}

\section {Star formation in blue compact and star burst galaxies}

In this section we investigate the correlations between the CO 
luminosities and the absolute magnitudes, H$_{\beta}$\ luminosities, masses
of
neutral gas, far infrared (FIR) luminosities and the 
temperatures of dust in the blue compact and star burst galaxies in this sample.
For completeness we have included (where possible) both detections and 
upper limits. 
The CO luminosity in the observed 35$\arcsec$ region of each galaxy was
calculated assuming $L_{CO} = (\pi r^{2})I_{CO}$\ (K km s$^{-1}$\ pc$^{2}$)
where $r$ is the radius of the 35$\arcsec$ beam on the galaxy in pc. 
The true CO luminosities may be somewhat higher since the complete
galaxy was not included in the 35$\arcsec$ beam in every case. 

The luminosities
of the galaxies which were detected are given in Table 2 and the upper
limits are given in Table 3. In Table 2 we have also collated the
observations of CO in BCGs made by Sage et al. (1992) and those by
Taylor, Kobulnicky \& Skillman (priv. com.). In these tables we have included the
absolute magnitudes, the FIR luminosities and the
HI masses. 
For consistency the absolute magnitudes (in Tables
2 and 3) and the HI masses (in Table 2), were 
obtained from the LEDA compilation (Paturel et al. 1997). 
Similarly the FIR luminosities for all galaxies were obtained from
Salzer \& MacAlpine (1988) (these are co-added IRAS survey data)
where available and from the LEDA compilation
otherwise. The FIR colour in Tables 2 and 3 is defined as
log$(f_{100}/f_{60})$\ where $f_{100}$\ and $f_{60}$\ are the FIR flux
densities
at 100$\mu$m and 60$\mu$m respectively. These fluxes were also obtained from
Salzer \& MacAlpine (1988) where available and from the IRAS Point Source
Catalogue (Lonsdale et al. 1985) otherwise. The uncertainty in the
$f_{100}$\ and $f_{60}$\ flux densities is between 5\% and 15\%;
corresponding to a maximum error of 20\% in the ratio $f_{100}/f_{60}$.

In comparing the CO luminosities with HI masses and the FIR luminosities the 
differences in the apertures with which these observations were 
made should be borne in
mind. The CO observations were made with a 35$\arcsec$\ 
field-of-view and the CO luminosities given in Tables 2 and 3
correspond only to the parts of the galaxies that were actually observed. 
The FIR observations (from the IRAS satellite) were made with apertures
of 1.5$\arcmin$\ and 3.0$\arcmin$\ at 60$\mu$m and 100$\mu$m respectively,
and the FIR luminosities of the complete 
galaxies will have been observed in all cases. Similarly, the HI observations 
have been made with a field-of-view larger than 35$\arcsec$\ in most cases.

The H$_{\beta}$\ luminosities given  in Table 2 and 3 were obtained from 
the equivalent widths of the H$_{\beta}$\ lines and the B-magnitudes given by
Salzer, MacAlpine \& Boroson (1989) and Terlevich et al. (1991)
respectively. The equivalent width of the H$_{\beta}$\ line is 
obtained from spectroscopic observations made with a narrow slit and it is very 
likely that the slit was positioned on the brightest HII region in
each galaxy in which these could be resolved. In using the 
equivalent width obtained from a narrow slit observation to obtain the 
luminosity of the entire galaxy we are assuming that the H$_{\beta}$\ 
surface brightness of the
galaxy is similar to that of the region where the slit observations were
made. This is unlikely to be true for all galaxies, particularly the
non-compact galaxies. 
The H$_{\beta}$\ 
luminosities were corrected for extinction with the Galactic extinction
law and the reddening coefficients
given by Salzer, MacAlpine \& Boroson (1989) and Terlevich et al. (1991).

\subsection{CO--Absolute Magnitude correlation}

The absolute magnitude of a galaxy is an indicator of its total 
stellar population, including as it does both the young stars in a 
starburst and the underlying older stars. Also, through the mass luminosity
relation, the absolute magnitude is indicative of the total 
stellar mass of a galaxy. The correlation between the CO luminosity and the
absolute magnitude of the galaxies in these samples is shown in Fig.2a.
In this sample of 34 galaxies, 14 galaxies are (equal to or) less luminous
than M$_{abs}$ = $-$18 (BCGs
) and 20 are more luminous than M$_{abs}$ = $-$18 (giants). 
CO has been detected in only 5 giants i.e. $\sim$25\% and not in any of
the dwarfs. These
observations of a large sample of BCGs confirm previous observations that 
BCGs are deficient in CO gas. 

As already mentioned it is
possible that the 35$\arcsec$ beam of OSO does not detect the total CO 
content of some  galaxies as their optical sizes, judged from 
their D$_{25}$\ diameters (Table 2), are considerably larger than 35$\arcsec$.
To check whether the CO luminosities of some of the
galaxies in Table 2 may have been significantly under-estimated,
we have included the observations of Young et al. (1986) in Fig. 2a. These 
authors have mapped CO in their sample of galaxies. In Fig. 2a, the data of
Young et al.  and the data of Sage et al. (1992) and Taylor, Kobulnicky 
\& Skillman (priv. com.) are linearly correlated with $M_{abs}$
and the present data are consistent with this correlation.
A straight line fit to all data (both detections and upper limits, but
excluding the data for NGC1569 and NGC6822) has
the form;
\[log(L_{CO}) = -1.96\pm 0.83 + (-0.51\pm 0.04)\times M_{abs} \]
with a Spearman's rho correlation coefficient of -0.81 \\
(ASURV software of Isobe,
LaValley \& Feigelson STARLINK SUN/13.2, MUD/005) 
The two discrepant points are for NGC1569 and NGC6822 (Taylor, Kobulnicky 
\& Skillman priv. com.). The D$_{25}$\ diameters of these two galaxies are 
45$\arcsec$\ and 151$\arcsec$\ 
respectively and it is not clear why the CO luminosities of these 
galaxies are lower than the luminosities of other galaxies of comparable absolute
magnitude. 

This strong
correlation between the CO luminosities and the absolute magnitudes of
these galaxies
suggests that the CO content of a galaxy is either determined by
or depends on its total stellar content.
It is also possible that the CO -- M$_{abs}$\ correlation actually conceals 
the CO $-$
metallicity correlation via the relation log(O/H)+12 = $-$0.2M$_{abs}$+4.86
(Arimoto, Sofue \& Tsujimoto (1996), Roberts \& Haynes (1994), Wilson
(1995)). 
It is not possible to confirm this possibility as metallicity 
data for the galaxies in the current sample are not available.
The correlation in Fig. 2a. thus suggests that
the CO luminosity of a galaxy is a tracer of
either the total stellar mass or the metallicity of a galaxy.

The galaxies in this sample with detected CO are relatively red; the mean 
(B-V)$_{cor} = $\ 0.5. This is in disagreement with the observations of
Israel, Tacconi \& Baas (1995) whose observations of dwarf galaxies
suggests that CO is preferentially detected in blue (dwarf) galaxies. 

\subsection{CO -- H$_{\beta}$\ correlation}

The H$_{\beta}$\ luminosity of a galaxy is proportional to the number of
high mass stars in the galaxy, allowing for extinction and geometric
factors (like the covering factor). The star-formation rate of the high mass
stars is given by  (Condon 1992)
\[ [\frac{SFR(M\geq 5M_{\sun})}{M_{\sun} yr^{-1}}] \sim 6\times
10^{-42} \eta^{-1} L_{H_{\beta}} \]
where $L_{H_{\beta}}$\ is the dereddened H$_{\beta}$\ luminosity of a
galaxy (in erg s$^{-1}$) and $\eta$ is the covering factor.
These high mass stars will only be a few million years old and will still
be associated with the giant molecular clouds from which they
formed. The CO -- H$_{\beta}$\ correlation thus 
provides a link between the molecular clouds in a
galaxy and the population of the high mass stars which form from these
clouds.

In Fig. 2b the CO luminosity is shown as a function of the H$_{\beta}$\
luminosity. Both the CO detections and upper limits have been plotted
along with the data of Sage et al. (1992) and those of Taylor, Kobulnicky
\& Skillman (priv. com.). The H$_{\beta}$\ luminosities in Fig.2b
suggest a SFR of $\sim 2  \times 10^{-2}$\ M$_{\sun}$\ yr$^{-1}$\ to
$\sim 6$\ M$_{\sun}$\ yr$^{-1}$. These SFRs have been obtained for a covering
factor  $\eta =$ 1.0, this is unlikely to be the case for all galaxies
in this sample. 

A straight line fit to all data in Fig.2b. (i.e. both detections and
upper limits) has the form;
\[ log(L_{CO}) = (-14.02\pm 16.44) + (0.51\pm 0.39)\times
log(L_{H_{\beta}})  \]
with a Spearman's rho correlation coefficient of 0.16 i.e. there is no
correlation between the CO and H$_{\beta}$\ luminosities. These data
suggest that the SFR, as obtained from the H$_{\beta}$\ luminosity of these 
galaxies, is independent of the CO content of these galaxies.
It is possible that the assumption that the covering factor 
$\eta$=1.0 for all our galaxies is wrong and that it
may in fact be a function of the
SFR; i.e. as the SFR increases, the wind and the supernovae
will disperse the surrounding gas and decrease the covering factor. 

\subsection{CO-HI correlation}

In order to investigate the relation between the CO and 
atomic gas contents of blue compact and star burst galaxies, the CO luminosity is plotted 
as a function
of the HI mass in Fig. 2c. The L$_{CO}$/M$_{HI}$\ ratio in this sample
varies from $5 \times 10^{-1}$\ to $8 \times 10^{-4}$. This range in
the L$_{CO}$/M$_{HI}$\ ratio is equal to the 
range seen in galaxy type 1 to 7 (Sage 1993). 
At low atomic gas mass the luminosity of CO increases with the mass of
the atomic gas, but the rate of increase slows down as the
atomic gas mass increases. For masses of 
about 10$^{10}$\ $M_{\sun}$\ and higher 
there is no further increase in the luminosity of CO. This `observation' 
depends to some degree on the data for UM274, although there is evidence for
flattening from atomic gas mass of 10$^{9}$\ $M_{\sun}$\ and higher. 
The D$_{25}$\ diameter of UM274 is 75$\arcsec$\  and if the CO density
over the visible galaxy is assumed to be similar to that seen over the
observed 35$\arcsec$, then the CO luminosity of this galaxy will
increase by about a factor of four (indicated by the arrow in Fig.2c).
This will however not remove the flattening seen in Fig. 2c.
The HI data for this galaxy were obtained with a field-of-view of
4$\arcmin$\
(E--W) and 22$\arcmin$\ (N--S) (Bottinelli, Gougenheim \& Paturel
1982). If we assume that the total HI gas is within 4$\arcmin$\ and the CO
is cospatial with HI then (for a uniform CO density) the total CO
luminosity of this galaxy will be a factor of about 70 higher than the
observed luminosity. This will still not completely remove the
flattening in Fig.2c. 
It is unlikely that the CO luminosities of galaxies with HI mass
greater than 10$^{9}$\ $M_{\sun}$\ will be significantly higher than
those given in Table 2 for, although the atomic gas can extend to several 
times the optical diameter, the CO gas is likely to be clumped around
star-forming regions rather than be globally distributed like the atomic gas. 
The change in the CO luminosity with increasing
atomic mass could be due to following reasons:
\begin{itemize}
\item At high atomic gas mass a large number of molecular gas clouds are
formed in the galaxy and some clouds will be optically thick so
that the
total CO content of these clouds will not be detected. Also, if optically
thick clouds shadow optically thin ones in both spatial and
velocity space then a fraction of the CO in a galaxy will not 
be detected. 
High resolution spatial and velocity maps of these galaxies in
molecular and atomic lines are required to explore this further.
\item Another possibility for the relation in Fig. 2c 
may be that the star-formation in UM274 and other
high HI mass galaxies in this study is independent of total
atomic gas mass. The star-formation  may be either a
stochastic process or be triggered by events external to
the galaxy e.g. by
collision(s) with other galaxies or intergalactic clouds. For
example, Turner,
Beck \& Hurt (1997) have suggested that the low CO content of NGC 5253
may be a consequence of accretion of low metallicity
intergalactic gas by this galaxy.
In the case of UM274, the `tails' in its optical image suggest 
a possible collision which may have triggered a highly efficient 
star-formation process and increased its atomic gas mass.
\end{itemize}

\subsection{CO-FIR correlation} 

The correlation between the CO luminosity and the FIR luminosity in BCGs
has been investigated by Rickard \& Harvey (1984), Young et al. (1984),
Sanders \& Mirabel (1985), Young et al. (1986) and others. Young et al.
(1986) found that the  CO luminosity is linearly correlated with FIR
luminosity
and that this correlation depends on the dust temperature. 
The correlation is particularly tight for galaxies in well 
defined temperature ranges.  The correlation between the CO 
luminosities and FIR luminosities of the current sample is 
shown in Fig 2d. As this sample is not large enough to divide 
the galaxies into dust temperature ranges, a power law was fitted to the 
all detections (open and filled circles) in Fig. 2d.  We find a relation of 
the form \[log(L_{FIR}) = (3.31\pm 0.84) + (0.79\pm 0.11)log(L_{CO}) \]
The Spearman's rho correlation coefficient for this fit is 0.88.
This result is consistent with the analysis of Young et al. 
(1986) and it would be tempting to conclude (as has been done 
by Young et al.  as well as by Tacconi \& Young 1987) that the CO 
luminosities of  galaxies increase linearly with their
FIR luminosities. However, {\em this would be
misleading} because the upper limits have not been included in the
linear fit. If the upper limits are included the Spearman's rho
correlation coefficient drops to 0.54 implying a weaker 
correlation.
There are large numbers of FIR-bright galaxies which are not 
CO bright and {\em the FIR luminosity of a galaxy is not a good tracer of the 
observed CO gas in a galaxy}. The lower CO luminosity of some FIR 
luminous galaxies may be due to the presence of optically 
thick clouds which cause shadowing.
This would be consistent with the CO--HI relation discussed in
section 3.3.

Since massive stars are formed in dusty giant molecular clouds, the FIR
luminosity is from dust heated by stars more massive than $\sim$5
M$_{\sun}$\ (Devereux \& Young 1990). The star-formation rate
is given by (Condon 1992)
\[ [\frac{SFR(M\geq 5M_{\sun})}{M_{\sun} yr^{-1}}] \sim 9\times
10^{-11} \eta^{-1} (L_{FIR}/L_{\sun}) \]
The SFR obtained from the FIR luminosity will be more accurate as
it will not have  been affected by extinction.
The SFR in these BCGs, for the FIR luminosities in Fig. 2d, range from
$\sim 3\times 10^{-4}$\ M$_{\sun}$\ yr$^{-1}$\ to $179$\ M$_{\sun}$\
yr$^{-1}$\ for a dust covering factor $\eta = 1$ . 

At low FIR luminosities the SFR is considerably
lower than that obtained from H$_{\beta}$\ luminosities.
However, at these low FIR luminosities the dust 
mass is about 50 M$_{\sun}$\ (for a dust temperature of 45 K
see section 3.5) and it is possible that the dust covering factor is not 
as high as that assumed here (the HI covering factor could still be high).

At high FIR luminosity the SFR derived from the FIR luminosities
is almost 30$\times$\ higher than that  computed from the
H$_{\beta}$\ luminosities, and this may be due to extinction of the latter. 
At these high FIR luminosities the dust mass increases to $\sim
10^{7}$\ M$_{\sun}$\  and the dust extinction would be 
expected to be greater. Also the 
H$_{\beta}$\ luminosity in section 3.2 has been 
corrected for reddening assuming a Galactic reddening law, but it is possible 
that in low metallicity galaxies the form of
the reddening law is different, e.g. steeper, like that observed for the 
LMC and SMC (Nandy et al. 1981,1982). The H$_{\beta}$\ luminosity will be
considerably higher if a correction with a steep
reddening law is applied.

\begin{figure*}[!ht]
\psfig{figure=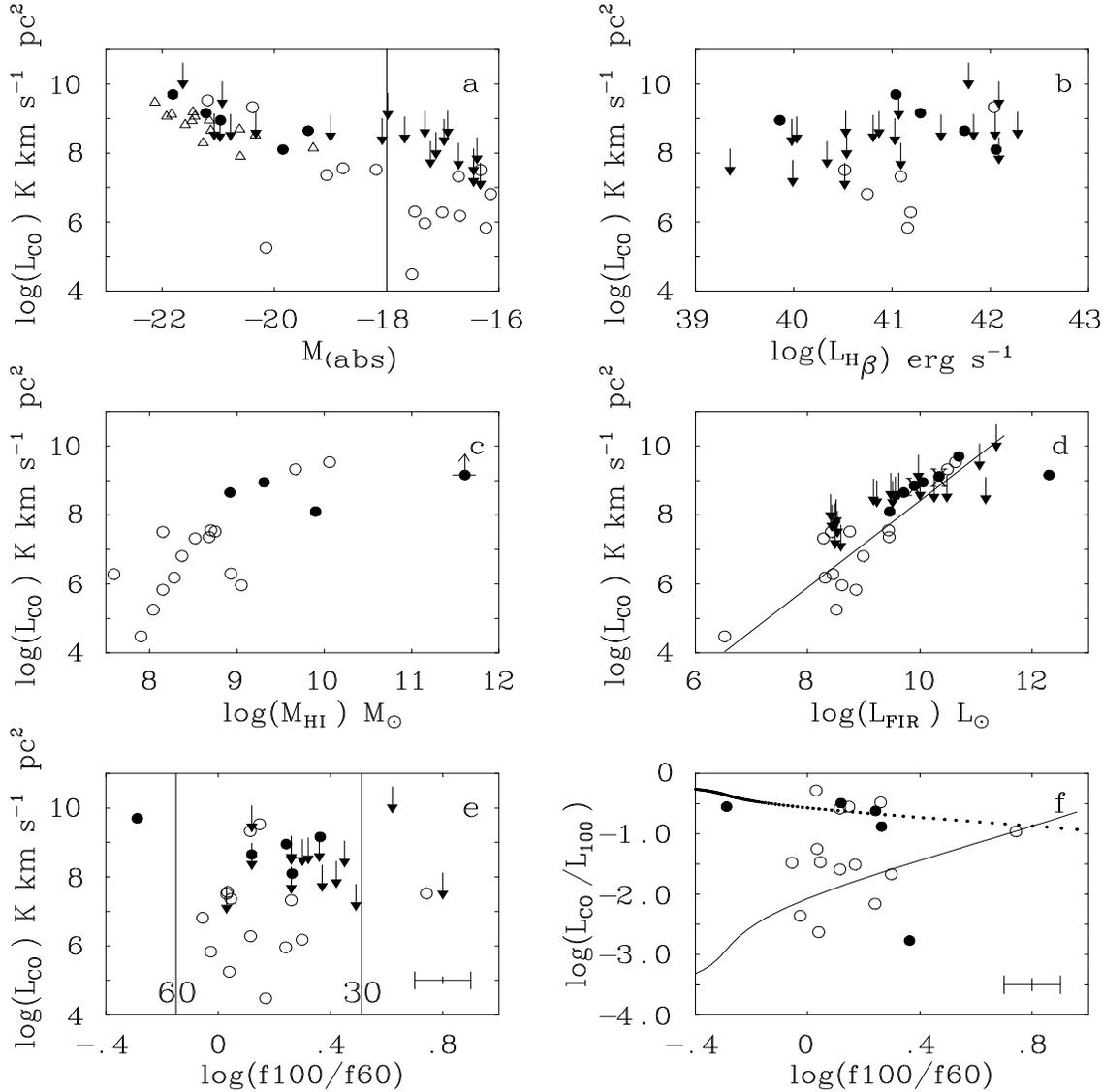,width=15.0cm,height=15.0cm}
\caption[ ]{The correlations with CO 
luminosity. In these figures the filled circles are the current observations 
and the open circles are those of Sage et al. (1992) and 
Taylor, Kobulnicky \& Skillman (priv. com.). In (a) the open triangles are 
data from Young et al. (1986).  
The panels show the following:
(a) the correlation with absolute magnitude,
(b) the correlation with  H$_{\beta}$\ luminosity,
(c) the correlation with HI mass. A typical error for
the HI mass is shown by one data point (UM274) at
top right. The arrow (for UM274) indicates the factor 4 increase in the 
CO luminosity if the complete optical galaxy had been 
observed for CO, (d) the correlation
with FIR luminosity - the straight line is a power law fit to the observed
data points and is described in section 3.4, (e) the correlation 
with FIR colour. Here the two lines denote the colours at 30 K 
and 60 K
 (emissivity of $\nu^{+1}$) respectively, (f) the temperature 
dependence of the L$_{CO}$/L$_{100}$\ ratio. The full line 
denotes a T$^{-4}$\ relation and the `string-of-beads'
denotes a T$^{+1}$\ relation. The 1$\sigma$\ error in
the luminosities is of the order of the size of the open and filled
circles. The 1$\sigma$\ error in the infrared colours is shown in the
bottom right corner of figures (e) and (f).}
\end{figure*}

\subsection{Temperature dependence of CO luminosity}

In Fig. 2e the CO luminosity is plotted as a function of the FIR colour.
The FIR colour is proportional to the temperature of the dust.
No correlation is seen between these two parameters. However, if
the dust and the CO gas are assumed to be in thermal equilibrium then
this figure suggests that CO  exists preferentially in clouds with dust
temperature between
30 K and 60 K (assuming dust emissivity of $\nu^{+1}$)
However, the distribution of data in Fig. 2e. almost certainly is due to a 
selection effect -- the dust temperature in the selected galaxies happens
to be between 30 K and 60 K. It is also possible that the 
(high temperature) dust detected at 60$\mu$m and 100$\mu$m 
and CO do not coexist.

To explore further this (lack) of correlation between CO luminosity and dust
temperature we have examined the dependence of the
L$_{CO}$/L$_{100}$\ ratio on dust temperature. 
Young et al. (1986) have shown that
if the CO gas  and dust are assumed to be in thermal equilibrium then
\[ L_{CO}/L_{100} \propto T_{d}^{-4}   \]

where $L_{100}$\ is the luminosity at 100$\mu$m. In this relation 
dust emissivity proportional to $ \nu^{+1}$\ is assumed. 
In Fig. 2f the ratio $L_{CO}/L_{100}$\ is plotted as a function of FIR
colour and the full line is the $T_{d}^{-4}$\ relation. 
A statistical fit of the
relation to these data has not been attempted as the assumptions in
its derivation and the errors in the observations do not warrant
such detail at present. The data in Fig. 2f suggest that for 
about 50\% of the galaxies in this sample the $L_{CO}/L_{100}$\ ratio is 
consistent with the $T_{d}^{-4}$\ relation, in agreement with the
conclusions of Young et al. . However, there
are equal number of galaxies in Fig. 4f for which $T_{d}^{+1}$\ more
accurately defines the relation between $L_{CO}/L_{100}$\ and the FIR
colour. If the molecular gas is in thermal equilibrium with the 
dust then this suggests a dust emissivity proportional to  $\nu^{-2}$\ 
in a large  fraction of galaxies in this sample.

\section{Conclusions}

We have attempted to observe $^{12}$CO in a sample 34 galaxies selected for
their high star-formation rate and a large spread in absolute luminosity
(or equivalently a large range in metallicity). This sample has 18 BCGs.
In this sample CO has been detected in only 5 galaxies and none of these
are BCGs. We have combined these observations with published
observations of CO to investigate star formation in these galaxies. The 
following conclusions have been reached:

\begin{enumerate}
\item  We have shown
that CO is difficult to observe, or deficient in BCGs 
less luminous than  $M_{abs} = -20$. Since absolute magnitude is correlated 
with metallicity, these observations confirm that CO is deficient 
in low metallicity galaxies.
\item The star-formation rate obtained from the H$_{\beta}$\ luminosity
of a galaxy is lower than that obtained from the FIR luminosity. This may
be due to the following possibilities (a) that the reddening 
in low metallicity galaxies is considerably steeper than the 
Galactic reddening law used in this analysis, or (b) that the
covering factor of the neutral hydrogen gas in these galaxies is considerably
different from the covering factor of dust in these galaxies.
\item The correlation between the CO luminosity and the FIR
luminosity of the BCGs is rather weak i.e. FIR luminosity is not a good
tracer of CO in a galaxy. 
\item In some galaxies the frequently
assumed $\nu^{+1}$\ dust emissivity may not apply and in these galaxies a
dust emissivity of $\nu^{-2}$\ may be a more satisfactory
explanation.
\end{enumerate}

\begin{acknowledgements}
This paper was produced with facilities provided by the STARLINK Project,
funded by PPARC at RAL. We would  like to acknowledgements the use of
LEDA extragalactic database. We are grateful to the TAC of OSO for
allocating observing time. EB acknowledges support from CONACyT via grant
number 0460P--E. 

\end{acknowledgements}

\end {document}